\documentclass[aps,prl,showpacs,twocolumn,superscriptaddress]{revtex4}
\usepackage{psfrag,graphicx}

\begin{document}

\title{Spintronics of a Nanoelectromechanical Shuttle}



\author{D. Fedorets\footnote{Present address: Kavli Institute of Nanoscience,
TU Delft, Lorentzweg 1, 2628 CJ Delft, The Netherlands.}}
\affiliation{Department of Physics,
G\"oteborg University, SE-412 96 G\"oteborg, Sweden}
\author{L. Y. Gorelik}
\affiliation{Department of Applied Physics, Chalmers University of
  Technology, SE-412 96 G\"oteborg, Sweden}
\author{R. I. Shekhter}
\affiliation{Department of Physics, G\"oteborg University, SE-412
96 G\"oteborg, Sweden}
\author{M. Jonson}
\affiliation{Department of Physics, G\"oteborg University, SE-412
96 G\"oteborg, Sweden}




\date{\today}
\begin{abstract}
We consider effects of the spin degree of freedom on the
nanomechanics of a single-electron transistor (SET) containing a
nanometer-sized metallic cluster suspended between two magnetic
leads. It is shown that in such a nanoelectromechanical SET
(NEM-SET) the onset of an electromechanical instability leading to
cluster vibrations and ``shuttle" transport of electrons between
the leads can be controlled by an external magnetic field.
Different stable regimes of this spintronic NEM-SET operation are
analyzed. Two different scenarios for the onset of shuttle
vibrations are found.
\end{abstract}




\maketitle




As the downsizing of electronic devices reaches the near molecular
scale, the Coulomb forces associated with inhomogeneous charge
distributions produced during device operation become comparable
with the chemical forces that hold the device together.
Consequently, the spatial configuration of a device may well
change dynamically during its operation. This inherent feature of
nanoelectronics can be turned into an advantage by designing the
devices with the mechanical degrees of freedom in mind; this is
the scope of nanoelectromechanics and the basis for
nanoelectromechanical systems (NEMS) \cite{Roukes01}.

A pronounced nanoelectromechanical phenomenon --- electron
transport by means of a nanoelectromechanical "shuttle" mechanism
--- has recently been predicted to occur in certain NEMS systems
as a result of a bias voltage-induced nanoelectromechanical
instability \cite{Gorelik98}. The most suitable device for the
experimental observation of this phenomenon is a
nanoelectromechanical single-electron transistor (NEM-SET). A
NEM-SET is a single-electron transistor (SET) with a movable
central island whose center-of-mass motion is confined by some
potential. Experimental studies of NEM-SET devices can be found in
Refs.~\cite{Park00,Erbe01,Scheible04}. They have also been
extensively investigated theoretically
\cite{Shekhter02,Fedorets03,Fedorets04,Novotny03,Smirnov04}.

 Another rapidly developing branch of condensed
matter physics is spintronics \cite{Zutic04}, which deals with
devices whose functionality depends on the control and
manipulation of the spin rather than the charge of electrons. A
recent development of great interest in this context is the
demonstration of a magnetic hybrid nanostructure consisting of a
single $C_{60}$-molecule placed between two ferromagnetic
electrodes \cite{Pasupathy04}. The possibility to manipulate the
spin of mechanically shuttled electrons in a molecular NEM-SET of
this kind brings about an exciting opportunity to trigger
nanomechanical vibrations in the THz range by means of a weak
external magnetic field much smaller than the magnetic anisotropy
fields in the leads. This Letter is devoted to exploring this very
phenomenon. We will show that the spin-dependent tunneling of
electrons between differently polarized leads results in a strong
sensitivity to an external magnetic field of the
nanoelectromechanical instability that is responsible for the
shuttling. Depending on the magnitude of the magnetic field two
different scenarios for the onset of nanoelectromechanical
shuttling are possible when the electrical field between the leads
reaches its critical value. This demonstrates the interesting
possibility to develop spintronics in the context of NEMS devices.

We will consider a NEM-SET with fully spin-polarized magnetic
leads \cite{Tokura00}. All electrons in the left lead are assumed
to have spins pointing up while in the right lead all spins are
pointing down. The movable central island has a single electron
energy level, which is spin-degenerate in the absence of a
magnetic field. A symmetric bias voltage is applied, resulting in
an electric field ${\cal E}$ between the leads. The external
magnetic field $B$ is oriented perpendicular to the direction of
the magnetization in the leads. We consider a symmetric coupling
to the leads, in which case the spin polarization in the leads
does not contribute to the magnetic field on the island.

 The Hamiltonian used to describe our system,
\begin{eqnarray}\label{Hamiltonian}
    {H} &=& \sum_{\alpha, k} \epsilon_{\alpha k} a_{\alpha k}^{\dag}
a_{\alpha k} +\sum_{\alpha, k} T_\alpha(X) \left[a_{\alpha
k}^{\dag} c_{\alpha} + c_{\alpha}^{\dag}a_{\alpha k} \right]
\nonumber\\&+& \left[\epsilon_0-e{\cal E}X\right] \sum_\alpha
c_{\alpha}^{\dag} c_{\alpha} - ({g\mu_B
B}/{2})\left[c_{\uparrow}^{\dag}
c_{\downarrow}+c_{\downarrow}^{\dag} c_{\uparrow}\right]
\nonumber\\&+&
Uc_{\uparrow}^{\dag}c_{\uparrow}c_{\downarrow}^{\dag}c_{\downarrow}
+ H_{osc}\,
\end{eqnarray}
has several terms. The first describes noninteracting electrons in
leads ($\alpha=L,R$), whose electron densities of states ${\cal
D}$ are assumed to be energy independent. The operator $a_{\alpha
k}^{\dag}$($a_{\alpha k}$) creates (destroys) an electron with
momentum $k$ in the lead $\alpha$ with the corresponding spin. The
electrons in each lead are held at a constant electrochemical
potential $\mu_{L,R} = \mp eV/2$, where $e<0$ is the electron
charge and $V>0$ is the bias voltage. Since the leads are fully
spin-polarized the lead index $\alpha$ can also be used as a spin
index: $L=\uparrow$ and $R=\downarrow$. The second term represents
tunneling of electrons (without spin flip) between the island and
the leads. The operator $c_{\alpha}^{\dag}$($c_{\alpha}$) creates
(destroys) an electron
 with spin $\alpha$ in the dot.
 The tunneling amplitudes depend exponentially on the displacement $X$:
 $T_{L,R}(X)=T_0 \exp\{\mp
 X/\lambda\}$.
 The third and fourth terms describe the single electronic state in the dot and
 its coupling to the electric field ${\cal E}$ and the magnetic field $B$.
 The Zeeman splitting is given by
$g\mu_B B$, where $g$ is the electronic $g$-factor and $\mu_B$ is
the Bohr magneton. The fifth term represents the Coulomb repulsion
between two electrons on the island and the last term $H_{osc} =
P^2/2M + M\omega_0^2X^2/2$
 describes the
vibrational degree of freedom associated with the center-of-mass
motion of the island. Here $M$ is the mass of the island and
$\omega_0$ its vibration frequency.
 For simplicity, we assume that the temperature is zero.

%

In the absence of an external magnetic field, electronic tunneling
through the system described by the
Hamiltonian~(\ref{Hamiltonian}) is blocked. Applying a bias
voltage and a magnetic field allows the electronic transport to be
externally manipulated by lifting this ``spin-blockade". By
deblocking electron tunneling the mechanical degree of freedom is
also greatly influenced, the equilibrium position of the island
becoming unstable if the rate of energy transfer from the
electronic subsystem to the nanooscillator exceeds a critical
value.

The coupled electronic and mechanical dynamics of the dot is
governed by a quantum master equation for the corresponding
reduced density operator $\rho(t)$. It can be derived from the
Liouville-von Neumann equation for the total system by projecting
out the degrees of freedom associated with the leads and the
thermal bath. The reduced density operator $\rho(t)$ obtained in
this way acts on the Hilbert space of the dot, which is a tensor
product of the Hilbert space of the oscillator and the electronic
space of the dot. The latter is spanned by the four basis vectors
$|0\rangle$, $|\uparrow\rangle\equiv c_\uparrow^\dag|0\rangle$,
$|\downarrow\rangle\equiv c_\downarrow^\dag|0\rangle$, and
$|2\rangle \equiv c_\downarrow^\dag c_\uparrow^\dag|0\rangle$. In
the electronic basis the operator $\rho(t)$ can be written as a
$4\times 4$ matrix whose elements are operators in vibration
space. The diagonal elements $\rho_{0}\equiv \langle
0|\rho|0\rangle$ and $\rho_{2}\equiv \langle 2|\rho|2\rangle$
represent the density operators of the empty and doubly occupied
oscillator correspondingly. The singly occupied oscillator is
described by the $2\times 2$ block $\hat{\rho}_1 \equiv
(\rho)_{s_1, s_2} \equiv (\langle s_1|\rho|s_2\rangle)$, where
$s_1,s_2=|\uparrow\rangle,|\downarrow\rangle$.

In the high bias-voltage limit \textbf{($eV>U>>\hbar\omega$)}
\cite{high:voltage} the time evolution of the density operators
$\rho_0$, $\hat{\rho}_1$ and $\rho_2$ is determined by the coupled
system of dimensionless equations of motion
\begin{eqnarray}\label{S:EOM:rho:0}
\partial_t {\rho}_{0} &=&
-i\left[H_{osc} +xd, \rho_{0} \right] - \left\{ {\Gamma}_{L}(x)
   , \rho_{0} \right\}/2
      \nonumber\\ &+&  {\rm Tr}_s \sqrt{\hat{\Gamma}_{R}^{\downarrow}}
\hat{\rho}_1 \sqrt{\hat{\Gamma}_{R}^{\downarrow}} + {\cal
L}_{\gamma} \rho_{0}\,,\\
 \label{S:EOM:rho:1}
\partial_t \hat{\rho}_1 &=&  -i\left[H_{osc}-\frac{h}{2}\hat{\sigma}_x,
\hat{\rho}_1 \right] - \left\{ \hat{\Gamma}_{+}^{\downarrow},
\hat{\rho}_1 \right\}/2 \nonumber\\&+&
 \sqrt{\hat{\Gamma}_{L}^\uparrow}\rho_{0}\sqrt{\hat{\Gamma}_{L}^\uparrow} +
 \sqrt{\hat{\Gamma}_{R}^\uparrow}
 \rho_{2}\sqrt{\hat{\Gamma}_{R}^\uparrow} + {\cal L}_{\gamma}
 \hat{\rho}_1\,,\\
\label{S:EOM:rho:2} \partial_t \rho_{2} &=& -i\left[H_{osc}-xd,
\rho_{2} \right] - \left\{\Gamma_R(x),\rho_{2}\right\}/2
  \nonumber\\ &+&
{\rm Tr}_s
\sqrt{\hat{\Gamma}_{L}^{\downarrow}}\hat{\rho}_1\sqrt{\hat{\Gamma}_{L}^{\downarrow}}
+ {\cal L}_{\gamma} \rho_{2}\,.
\end{eqnarray}
Here all lengths are measured in units of the zero point
oscillation amplitude $x_{0}$, all energies in units of $\hbar
\omega_0$ and time in units of $\omega_0^{-1}$; $x\equiv X/x_0$
and $p\equiv x_0 P/\hbar$ are dimensionless operators for the
oscillator displacement and momentum correspondingly.
Dimensionless electric ($d$) and magnetic ($h$) fields are defined
by
\begin{equation}\label{fields}
d \equiv e{\cal E}/(M\omega_0^2 x_0)\,,\quad h \equiv g\mu_B B /
(\hbar \omega_0)\,.
\end{equation}
The tunneling of electrons is described by the dimensionless
parameters $\Gamma_\alpha(x) \equiv 2\pi {\cal D}T_\alpha^2(x+d)/
(\hbar \omega_0)$, $(\hat{\Gamma}_\alpha^s)_{s_1, s_2} \equiv
\Gamma_\alpha(x) \delta_{s_1, s} \delta_{s_2, s}$ and
$\hat{\Gamma}_{+}^s \equiv \hat{\Gamma}_L^s + \hat{\Gamma}_R^s$.
The damping of vibrations is introduced via the simplest form of
the damping Liouvillian,
\begin{equation}\label{L:gamma}
 {\cal L}_{\gamma} \bullet
\equiv -i\gamma\left[x,\left\{p,\bullet\right\}\right]/2 -\gamma
(n_{\omega_0}+1/2) \left[x,\left[x,\bullet\right]\right]/2\,,
\end{equation}
 where $\gamma\ll 1$ is a
dimensionless dissipation rate and $n_{\omega_0}\equiv
1/[e^{\beta\hbar\omega_0}-1]$ is the Bose distribution function.
One can formally derive Eq.~(\ref{L:gamma}) by weakly connecting
the oscillator to
  an Ohmic heat bath by adding the terms $H_{bath}=\sum_q \hbar\omega_q b^{\dag}_q
b_q$ and $H_{osc-bath}=X\sum_q g_q (b^{\dag}_q+b_q)$ to the
Hamiltonian (\ref{Hamiltonian}) and then eliminate the bath
variables in the Born-Markov approximation \cite{Weiss99}.


The mechanical degree of freedom alone is described by the density
operator $\rho_{+}\equiv \rho_0 + {\rm Tr}_s \hat{\rho}_1 +
\rho_2$, which is conveniently analyzed in the Wigner
representation defined by
\begin{eqnarray}
  W_{\rho}(x,p) \equiv \frac{1}{2\pi} \int_{-\infty}^{+\infty} d\xi e^{-ip\xi}
  \left\langle x+\xi/2 \left| \rho \right|
  x-\xi/2\right>\,
\end{eqnarray}
for a density operator $\rho$.

In an experimentally relevant regime \cite{Park00}, where the
electromechanical coupling parameter
\begin{eqnarray}
\eta \equiv d/\lambda \sim 1/\lambda \ll 1
\end{eqnarray}
 is small, one can find the
stationary solution of the system of
Eqs.~(\ref{S:EOM:rho:0},\ref{S:EOM:rho:1},\ref{S:EOM:rho:2})
perturbatively in terms of the small parameters $\eta$ and
$\gamma$.

Rescaling the phase variables $X \equiv x/\lambda$, $P \equiv
p/\lambda$ and changing to the polar coordinates $X=A\sin\varphi$,
$P=A\cos\varphi$, one obtains the steady-state equation for the
Wigner function $W_{+}\equiv W_{\rho_+}$ of the oscillator as
\begin{eqnarray}
\label{S:steadyEOM:W:plus}
\partial_{\varphi}{W}_{+}  &=& \left\{ -
{\Gamma}_{+}(X){\cal C}/2 + {\cal L}_\gamma\right\}W_{+}
\\
&+&\left\{\eta \partial_P   + {\Gamma}_{-}(X){\cal C}/2
\right\}W_{0-2} - {\Gamma}_{+}(X){\cal
C}{W}_{\downarrow-\uparrow}/2 \,.\nonumber
\end{eqnarray}
Equation~(\ref{S:steadyEOM:W:plus}) couples to the steady state
equation for the vector-function $\textbf{W}\equiv
\left[W_{0-2},\,{W}_{\downarrow-\uparrow},\,W_{-c},\,W_{+c},\,W_{0+2}\right]^{T}$,
where $W_{0\pm 2} \equiv W_{\rho_0} \pm W_{\rho_2}$,
$W_{\downarrow-\uparrow} \equiv W_{\rho_{\downarrow \downarrow}} -
W_{\rho_{\uparrow\uparrow}}$ and $W_{\pm c} \equiv
W_{\rho_{\uparrow \downarrow}} \pm W_{\rho_{\downarrow
\uparrow}}$. Here $\Gamma_{\pm}(X)\equiv
\Gamma_R(X)\pm\Gamma_L(X)$, ${\cal L}_\gamma \equiv \gamma
\left[\partial_P P + (1/2\lambda^2)\partial_P^2\right]$ and ${\cal
C} \equiv \cosh \left[i\lambda^{-2}\partial_P\right]-1$.

One can show that in the leading order approximation the Wigner
function $W_{+}(A,\varphi)$ is $\varphi$-independent and
determined by
\begin{equation}\label{S:FP}
\partial_A A\left[ f(A) + D(A)\,\partial_A\right] {W}_{+}(A)
= 0 \,,
\end{equation}
where $f(A) = (A/2)\left[{\gamma} - \eta
\beta_0(A)\right]$, $
 A\beta_0(A) = - \int_{0}^{2\pi}
 (d\varphi/\pi)\,  \cos \varphi \,
 {G}_{0-2}^{(0)}(\varphi) \geq 0$
  and $D(A)>0$ is of second order in $\eta$ and $\gamma$.
The function ${G}_{0-2}^{(0)}(\varphi)$ is determined by the
system of differential equations:
\begin{eqnarray} \label{steadyEOM:G:0-2}
D_{\varphi}{G}_{0-2}^{(0)} &=& {\Gamma}_{-}(X)
{G}_{\downarrow-\uparrow}^{(0)}
 + {\Gamma}_{-}(X)\,, \\ \label{steadyEOM:G:down:up}
 D_{\varphi}{G}_{\downarrow-\uparrow}^{(0)} &=& {\Gamma}_{-}(X)
G_{0-2}^{(0)} -
 2h G_{-c}^{(0)} -
 {\Gamma}_{+}(X)\,, \\ \label{steadyEOM:G:c}
 D_{\varphi}{G}_{-c}^{(0)} &=& 2h
G_{\downarrow-\uparrow}^{(0)}\,,\quad D_{\varphi} \equiv
2\partial_\varphi +
 {\Gamma}_{+}(X)
\end{eqnarray}


It follows from Eq.~(\ref{S:FP}) that the Wigner function $W_{+}$
has the form $ W_{+}(A)\approx  {\cal Z}^{-1} \exp\left\{-
\int_{0}^{A}\, d A\,\frac{f(A)}{D(A)}\right\}$ and is peaked for
amplitudes $A_M$ determined by the conditions $f(A_M)=0$ and
$f'(A_{M})>0$. In the vicinity of these amplitudes ${W}_{+}(A)$
can be approximated by a narrow Gaussian of variance $\sigma^2
=D(A_{M})/f'(A_{M})$.
The behavior of the stationary solution $W_{+}(A)$ is determined
by the structure of the positive definite function $\beta_0(A)$,
which is bounded, has only one maximum and decreases monotonically
for large $A$. One can show that if $h< h_c$, the function
$\beta_0(A)$ has its only maximum at $A=0$, while if $h>h_c$ it
has a minimum there. In the vicinity of $A=0$, one finds that
$\beta_0(A)= \gamma_{thr}/\eta + {\cal O}(A^2)$, where
\begin{equation}
\gamma_{thr} \equiv \eta \frac{2\Gamma_0}{1+\Gamma_0^2}
\frac{h^2}{h^2+\Gamma_0^2}\,,\quad \Gamma_0 \equiv
\frac{1}{2}\Gamma_{+}(0)\,.
\end{equation}
From this equation it follows that the electromechanical
properties for a low-transparency junction (with a resistance in
the Gohm range) is sensitive to very weak magnetic fields of order
1-10 Oe. Such weak fields have a negligible effect on the internal
magnetization of the leads, which is why this effect was not
considered here.

\begin{figure}[htbp]
\begin{center}
    \includegraphics[width = 6.0 cm]{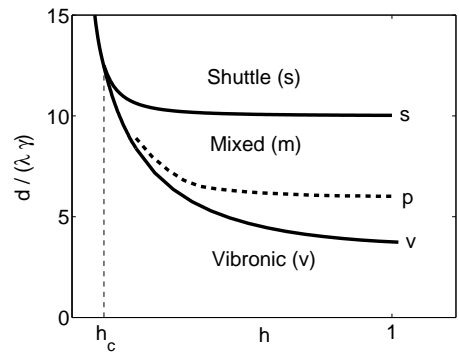}
    \caption{"Phase diagram" (for $\Gamma_{+}(0)=0.1$) in the electric field ($d$)- magnetic
    field ($h$) plane (see Eq.~(\ref{fields}))
     showing domains of different nanoelectromechanical behavior
    of the studied NEM-SET device. In the vibronic domain ($v$) only the
    vibrational ground state of the device is stable, in the shuttle
    domain ($s$) the stable state corresponds to developed island
    vibrations and in the mixed domain ($m$) both states are
    locally stable with probabilities that are equal along the dashed line $p$.
    }
   \label{S:phase}
   \end{center}
\end{figure}

The presence of various stationary regimes  can be illustrated by
a ``phase diagram'' in the $(d,h)$-plane. Figure~\ref{S:phase}
shows three domains that correspond to three different types of
behavior of the nanomechanical oscillator. In the ``vibronic''
domain ($v$), defined by the condition $\eta/\gamma < 1/\left[\max
\beta_0(A)\right](h)$, the NEM-SET system is stable with respect
to mechanical displacements of the island from its equilibrium
position. The ``shuttle'' domain ($s$) is determined by the
condition $\gamma < \gamma_{thr}$ and corresponds to developed
island vibrations of a classical nature. The third domain is the
``mixed'' domain ($m$). It appears because the $v$- and
$s$-regimes become unstable at different combinations of electric
($d$) and magnetic ($h$) fields if $h$ exceeds a critical value
$h_c$ (cf. \cite{Gorelik98c,Novotny03}).

Hence, while the shuttle regime is unstable below the line $v$ in
Fig.~\ref{S:phase}, the vibronic regime becomes unstable above the
line $s$. Between these lines (in the $m$-domain) both states can
be stable. The oscillator ``bounces'' between the $v-$ and
$s$-regimes due to random electric forces caused by stochastic
variations of the grain charge associated with tunneling events.

The transition time between the two locally stable regimes of the
$m$-domain is given by $\tau_{v \leftrightarrow s} = \omega_0^{-1}
\exp \left({S_{v \leftrightarrow s}}/\eta\right)$, where $S_{v
\leftrightarrow s} (d,h)\sim 1$. Since $S_{v \to s} \ne S_{s \to
v}$ and $\eta \ll 1$, the difference between the switching rates
$\tau_{v \to s}$ and $\tau_{v \leftarrow s}$ can be exponentially
large. This implies that the probabilities for the system to be in
one or the other of the two regimes can be very different. The
line $p$ in Fig.~\ref{S:phase} corresponds to $S_{v \to s} = S_{v
\leftarrow s}$ and therefore to equal rates for the transitions $v
\to s$ and $s \to v$. Below this line, the probability for the
system to be in the $v$-regime is exponentially larger than for it
to be in the $s$-regime, while above the dominance of the
$s$-regime is exponentially large. Due to the smallness of the
electromechanical coupling, $\eta \ll 1$, the transition between
the two regimes is very sharp. Hence the change of vibration
regime can be regarded as a ``phase transition''. Such a
transition will manifest itself if the variation of the external
fields is adiabatic on the time scale of $\max\{\tau_{s
\leftrightarrow v}\}$. One can expect enhanced low-frequency
noise, $\omega \lesssim \tau_{s\leftrightarrow v}^{-1}$, around
the line $p$ as a hallmark of the transition.

In the opposite non-adiabatic limit, either the $s$- or the
$v$-regime is ``frozen in'' in the mixed domain after crossing the
line $p$. Thus if one starts in the $v$-domain the $v$-regime
persists until the system crosses the line $s$, and if one starts
from the $s$-domain the $s$-regime persists until  the system
crosses the line $v$. Hence, one observes a hysteretic behavior of
the non-adiabatic shuttle transition.

\begin{figure}[htbp]
\begin{center}
    \includegraphics[width = 6.0cm]{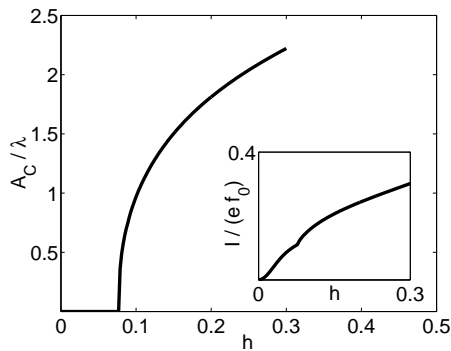}
    \caption{Steady-state amplitude $A_C$ of shuttle vibrations and current $I$
    through the NEM-SET device for parameters corresponding to a "soft" transition
    between the vibronic and shuttle domains of Fig.~\ref{S:phase}
    ($\Gamma_{+}(0)=0.1$ and $d / (\gamma\lambda) = 14.3$).
    }
   \label{S:limit:soft:1}
   \end{center}
\end{figure}

\begin{figure}[htbp]
\begin{center}
    \includegraphics[width = 6.0cm]{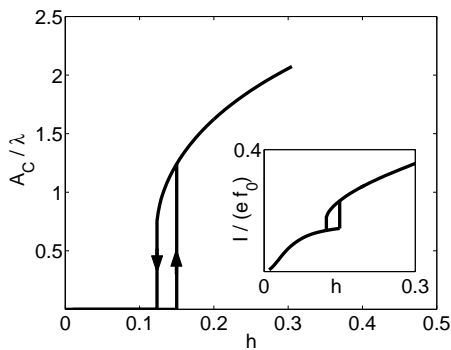}
    \caption{Hysteretic behavior of the steady-state shuttling amplitude $A_C$
    and current $I$ through the NEM-SET device for parameters
    corresponding to a "hard" transition between the vibronic and shuttle
    domains of Fig.~\ref{S:phase} ($\Gamma_{+}(0) = 0.1$ and
    $d / (\gamma\lambda) = 11.1$).
    }
   \label{S:limit:hys:1}
   \end{center}
\end{figure}

As one can see from Fig.~\ref{S:phase}, there are two different
scenarios for the onset of shuttle vibrations. If one crosses over
from the $v$- to the $s$-domain when $h < h_c$, i.e. avoiding the
mixed domain, the onset is \textit{soft}. In this scenario the
vibration amplitude grows gradually from zero to some finite value
after crossing the border line (Fig.~\ref{S:limit:soft:1}). If $h
> h_c$, on the other hand, the onset is \textit{hard}. In this case
the vibration amplitude shows a step at the transition point
(Fig.~\ref{S:limit:hys:1}), which corresponds to crossing either
the $p$- or the $s$-line depending on whether the transition is
adiabatic or not.

One can show that the expression for the steady state current
through the system is
\begin{eqnarray}
I = {e}\int\int dX dP\,
\Gamma_L(X)[W_{+}+W_{0-2}+W_{\downarrow-\uparrow}]/2\,.
\end{eqnarray}
Typical plots of the current in the cases of soft and hard
transitions are shown in Figs.~\ref{S:limit:soft:1} and
\ref{S:limit:hys:1}, respectively.

In conclusion, we have considered "shuttle" phenomena in a
nanoelectromechanical single-electron transistor (NEM-SET) system
consisting of a metallic island suspended between spin-polarized
leads. We have shown, that a coupling between the transport of
spin-polarized electrons and the center-of-mass motion of the
island allows us to control the dynamics of the mechanical degree
of freedom of the island by an external magnetic field. Different
stable operating regimes of the magnetic NEM-SET were found and
transitions between them induced by varying the electric and
magnetic fields were analyzed. We have hence demonstrated that
magnetic-field-controlled spin effects can lead to a very rich
behavior  of nanomechanical systems.

Although we have considered fully spin-polarized leads and assumed
a symmetric set-up, the overall qualitative picture does not
change if these conditions are relaxed. A partial polarization,
allowing a shuttle instability even at zero external magnetic
field, makes the system sensitive to the magnetic field in a more
narrow interval of electric fields. An asymmetric coupling to the
leads induces an additional magnetic field along the direction of
their spin-polarization. This field reduces the rate of energy
pumping into the oscillator, moving the domains in
Fig.~\ref{S:phase} to higher values of electric and magnetic
fields.

This work was supported in parts by the Swedish Foundation for
Strategic Research, by the Swedish Research Council and by the
European Commission through project FP6-003673 CANEL of the IST
Priority. The views expressed in this publication are those of the
authors and do not necessarily reflect the official European
Commission's view on the subject.


\begin{thebibliography}
\bibitem{high:voltage} Voltages
at which the shuttle phenomenon should occur in realistic systems
are large compared to the Coulomb energy $U\sim 1$~eV
\cite{Pasupathy04}.
\end{thebibliography}
\end{document}